\documentclass[11pt,twoside]{acta}
\SetPages{1}{16}
\SetVol{55}{2005}

\begin{document}
\begin{Titlepage}

\Title { Mass and radius determination for the neutron star in
   X-ray burst source 4U/MXB 1728-34}
\Author
{A. Majczyna \and J. Madej}
{Warsaw University Observatory, Al. Ujazdowskie 4, 00-478 Warsaw, Poland \\
email: jm@astrouw.edu.pl}
\end{Titlepage}

\abstract{We analyzed archival X-ray spectra of MXB 1728-34 obtained in 
1996-99 by the Proportional Counter Array on board of the RXTE satellite. 
X-ray spectra were fitted to our extensive grids of model atmosphere
spectra to determine the effective temperature $T_{\rm eff}$ on the neutron
star surface, logarithm of surface gravity $\log{g}$, and the gravitational
redshift $z$ simultaneously. We have chosen fitting by numerical model
spectra plus broad Gaussian line, modified by interstellar absorption and
the absorption on dust. We arbitrarily assumed either hydrogen-helium
chemical composition of a model atmosphere, or H-He-Fe mixture in solar
proportion. The statistically best values of $\log{g}$, and $z$ 
subsequently were used to determine mass and radius of the neutron star.
We obtained the best values of the parameters for the neutron star in
X-ray burst source MXB 1728-34: mass either $M=0.40$ or 
$0.63 M_\odot$ (for H-He or H-He-Fe models, respectively), radius  
$R=4.6$ or $5.3$ km, $\log g=14.6$ or $14.6$ and the gravitational redshift
$z=0.14$ or $0.22$. All the above parameters have very wide 1-$\sigma$
confidence limits. Their values strongly support the equation of state for
strange matter in MXB 1728-34.

{\bf Key words} {\it Stars: fundamental parameters -- stars: neutron -- X-rays:
  bursts -- stars: individual (MXB 1728-34) }

}


\section{Introduction}
X-ray source MXB 1728-34 was discovered in 1976 by Forman, Tananbaum
\&Jones (1976) in the survey observations of the Uhuru satellite. Type I
X-ray bursts from this source were identified in the same year by Lewin,
Clark \& Doty (1976) in SAS-3 satellite data. Optical counterpart still
remains unidentified probably due to large extinction in the direction to
this source. Recently Mart{\'\i} et al. (1998) has identified an infrared
counterpart  to MXB 1728-34, but the connection with the X-ray source has
not yet been confirmed. MXB 1728-34 was observed in a wide energy range
from radio  (Mart{\'\i} et al. 1998) to $\gamma$-ray energies (cf. Claret
et al. 1994). 

The determination of mass and radius of a neutron stars is a very important
and interesting problem, because it allows one to determine or to constrain
the equation of state of superdense matter. Distance to the source is also
necessary to localize the source relatively to the Galactic center.
For MXB 1728-34 the first estimate of its distance $d$ and true radius
$R$ was given by van Paradijs (1978), $d=4.2\pm 0.2$ kpc and $R=6.5\pm 0.4$
km (see also Foster et al. 1986). Both parameters were estimated assuming
that the peak flux of a given burst from this source approached the
Eddington limiting flux. It became immediately clear that such a radius $R$
was smaller than the minimum radius expected for a neutron star of the
canonical mass of $1.4 M_\odot$.

MXB 1728-34 was frequently observed in X-rays after that year,
and parameters of the neutron star were estimated in several papers.
Kaminker et al. (1989) interpreted spectra of this source using simplified 
emission models. They have derived the following values of the neutron
star parameters: mass $M=1.4-2.0\, M_\odot$ and radius $R=6.5-12$ km 
assuming the distance $d=6$ kpc and pure helium atmosphere.

Di Salvo et al. (2000) and Galloway et al. (2003) also assumed, that 
luminous bursts from this source are standard candles of the
bolometric luminosity at the Eddington limit. They concluded that the 
distance to this X-ray burster is in the range 4.4 -- 5.1 kpc.

On the other hand, Shaposhnikov et al. (2003) obtained other values:
$d=4.5-5.0$ kpc, $M=1.2-1.6 M_\odot$, and $R=8.7-9.7$ km. They also
concluded, that the helium mass abundance ${\rm Y} >0.9$ in atmosphere of
the neutron star in MXB 1728-34. These authors used semianalytical models
of expanding atmospheres by Titarchuk (1994), and Titarchuk \& Shaposhnikov
(2002). 

A different approach to this problem was used by Li et al. (1999),
who interpreted power spectra of the X-ray light curve and compared them
to the model by Osherovich \& Titarchuk (1999) and Titarchuk \& Osherovich
(1999). As the result Li et al. (1999) constrained the area on the
($M$, $R$) plane, where the neutron star of MXB 1728-34 is located. 
Constrained localization of MXB 1728-34 was characteristic for a star
built up of strange matter.  

We present here for the first time the estimation of mass and radius
of MXB 1728-34, obtained by fitting of archival RXTE X-ray spectra of
this source to the new extensive grids of model atmospheres and theoretical
spectra of neutron stars. This technique also allowed us to obtain the
upper limit of the distance to MXB 1728-34.
Theoretical model spectra were obtained with the {\sc ATM21}
code, which computes model atmospheres of hot neutron stars with the
account of Compton scattering on free electrons. The code takes into
account angle-averaged Compton scattering of X-ray photons with
initial energies approaching the electron rest mass.

Detailed description of the equations and numerical methods were given in a
long series of earlier papers (Madej 1991a,b; Joss \& Madej 2001; Majczyna
et al. 2002; Madej, Joss
\& R\'o\.za\'nska 2004; Majczyna et al. 2005).   
\clearpage
\section{Model atmospheres and theoretical spectra}

Model atmospheres used in this paper are based on the equation of 
radiative transfer of the following form
\begin{eqnarray}
\label{equ:tra}
& & \mu \, {\partial I_\nu (z,\mu) \over {\rho \, \partial z}}  = 
 \kappa^\prime_\nu (1-e^{-h \nu/kT})\, (B_\nu-I_\nu) \\
& + & \left( 1+{c^2 \over {2h \nu^3}} I_\nu \right)
 \oint \limits_{\omega^\prime } {d\omega^\prime \over 4\pi}
\int \limits _{0}^{\infty} {\nu \over {\nu^\prime }} \, \sigma (\nu^\prime
\rightarrow \nu , \vec n {^\prime} \cdot \vec n ) \, I_{\nu ^\prime}
(z, \vec n^\prime ) \, d\nu ^\prime \,   \nonumber  \\  &- &\,
I_\nu (z, \mu ) \oint \limits_{\omega ^\prime} {d\omega^\prime \over {4\pi}}
\int \limits _{0}^{\infty}  \sigma (\nu \rightarrow \nu ^\prime ,
\vec n \cdot \vec n {^\prime} ) \, (1+ {c^2 \over {2h{\nu ^\prime }^3 }}
I_{\nu ^\prime} ) \, d\nu ^\prime \, ,  \nonumber
\end{eqnarray}
which defines a very elaborate source function $S_\nu$ of our models 
(Madej 1991a). The significance of symbols in Eq. 1 is commonly
known and will not be repeated here. 

We stress here, that our theoretical models use sophisticated Compton scattering cross-sections 
$\sigma (\nu \rightarrow \nu ^\prime , \vec n \cdot \vec n {^\prime} )$,
which allow for a large photon energy change at the time of a single 
scattering off electrons with a relativistic thermal velocity distribution
(Guilbert 1981). Frequently used Kompaneets approximation was rejected
in our calculations in order to obtain high numerical accuracy of 
theoretical Comptonised X-ray spectra.

The actual code is able to compute model atmospheres and
theoretical X-rays spectra of very hot neutron stars, taking also
into account numerous bound-free and free-free monochromatic opacities of
various elements, and the equation of state of ideal gas. The code is 
``exact'' in that it solves the equation of transfer
coupled with the equation of radiative equilibrium using partial
linearisation and variable Eddington factors technique {(Mihalas 1978).}

We calculated 3 extensive grids of model atmospheres of hot neutron stars
corresponding to various arbitrarily assumed chemical compositions. 
\begin{itemize}
\item[--] Hydrogen-helium mixture with solar helium number abundance,
$N_{He}/N_H=0.11$ (223 models), 
\item[--] Hydrogen-helium-iron mixture with $N_{He}/N_H=0.11$ and solar
iron number abundance, $N_{Fe}/N_H=3.7 \times 10^{-5}$ (228 models), 
\item[--]  Hydrogen-helium-iron mixture with $N_{He}/N_H=0.11$ and
100 $\times$ solar
iron number abundance, $N_{Fe}/N_H=3.7 \times 10^{-3}$ (229 models).
\end{itemize}
Computed models cover the range of $1\times 10^7 \le T_{\rm{eff}}\le
3\times 10^7$ K with step of $10^6$ K, and the range of surface gravity
$15.0 \ge \log{g} \ge \log{g_{cr}}$ (cgs units) with step of 0.1. Here we
introduced the critical gravity $g_{cr}$, for which acceleration 
 exerted by the radiation pressure gradient and directed outward just
balances gravity. Finally, we transformed numerical ASCII models to FITS
format required by the {\sc xspec} package. 

\begin{figure}[!h]
\includegraphics[scale=0.5]{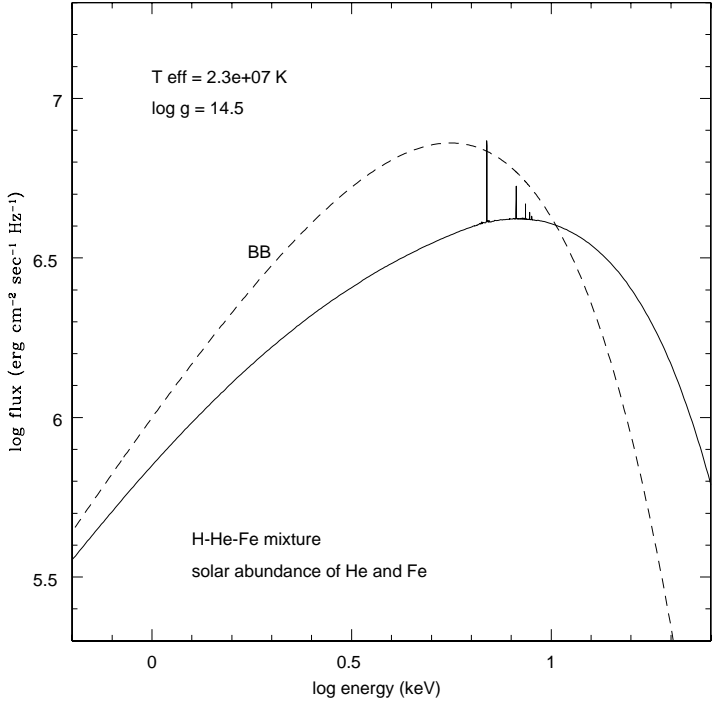}
\caption{{ \footnotesize Comparison of the neutron star theoretical spectrum
(solid line) and the blackbody spectrum (dashed line). Sample effective
temperature $T_{\rm eff}$, surface gravity $\log g$ and chemical
composition correspond to one of the best fits. The BB spectrum in this Figure was
computed for the temperature $T_{\rm eff}$.  }}
\label{fig:widma23}
\end{figure}

Properties of neutron star theoretical spectra were extensively discussed
e.g. in Majczyna et al. (2005), see also earlier theoretical papers of this
series. As an example, Fig.1 presents a model spectrum
with parameters similar to these of MXB 1728-34, see Section 6 (solid line).
The relevant effective temperature equals to
$T_{\rm eff}=2.3\times 10^7$ K, surface gravity $\log g=14.5$ (cgs), and
chemical composition is H-He-Fe of solar
proportions. Fig.1 includes also properly normalized
blackbody spectrum of the temperature $T=T_{\rm eff}=2.3\times 10^7$ K. 
Spectrum of blackbody was routinely used in many earlier papers for the 
rough representation of an X-ray burst spectrum.

Fig. 1 demonstrates basic properties of our model
spectrum. First, the latter is shifted to higher energies as compared with
the blackbody. Such a shift is typical for scattering atmospheres (Madej 1974)
and causes that the ratio of color to effective temperatures, $T_{\rm
col}/T_{\rm eff} > 1$. Second, one can note that the shape of precisely
computed spectrum is different than the blackbody and exhibits low energy
excess and several spectral lines. In this Figure spectral
lines are caused by hydrogenic iron and appear in emission. 

Both the specific shape of an exact spectrum and its shift toward higher
energy depend on $T_{\rm eff}$, $\log g$ and also the redshift $z$.
Therefore, these parameters can be determined by comparing of theoretical
and observed RXTE spectra. While RXTE spectral resolution does not allow to
resolve most of spectral lines, they contribute to theoretical counts in
channels of RXTE proportional counters.


\section{Observations and data analysis}
X-ray burster MXB 1728-34 was observed by the RXTE satellite  many times
at the time period 1996-1999. In our paper we have reanalyzed archival
RXTE spectral observations taken from the HEASARC database, numbered
10073-01-02-00, 10073-01-03-000, 20083-01-01-01, 20083-01-02-01,
20083-01-04-00, 20083-01-04-01. We have selected spectra obtained by  the PCA instrument, since it
was sensitive in the energy band 2-60 keV, and have chosen data from
the top layer of detectors Pcu0, 1, 2. The above spectra correspond to the
quiescent phase of the burster.

Standard-2 configuration was used to analyze the spectra. This type of
configuration has a very good energy resolution, however, it has poor time
resolution. We integrated raw spectra over  96 second intervals, taking
shorter spectra binned to 16-sec at the time of their extraction. We
selected spectra of this source outside bursts, because wanted to avoid
possible phases of radius expansion of the neutron star (NS), and
rapid changes of its luminosity. 
We did not include correction for dead time.  In our case, count rate
was not large and neglecting this correction generated error lower
than 1\%. 

During preparation of an observed  spectrum for fitting we neglected counts
in channels below 3.0 keV, since one can expect poor energy resolution
in low energy channels and too strong impact of interstellar absorption. 
Data above 20 keV were ignored due to poor statistics of counts. 
We extensively used the publicly available software {\sc xspec} v. 11.1 and
the response matrix v. 10.1. The {\sc xspec} software was described in
Arnaud (1996). 

Claret et al. (1994) have discovered a hard energy excess of this source in the 30--200 keV energy
band with the SIGMA telescope on board GRANAT satelite. Authors
fitted hard X-ray spectrum
of MXB 1728-34 by a thermal brem\-sstra\-hlung model with a very high electron temperature
of 38 keV = 4.4 $\times 10^8$ K. Authors did not explain what is the origin
of this extremely hot medium. The existence of very hot electrons there was
not confirmed at a later time.

RXTE observed the source MXB 1728-34 for a very long accumulated time period, and therefore
recorded many hard X-ray photons. As was written above, we integrated our spectra
over relatively short time period. For such integration time hard part of the spectrum is
very weak. For example, in our template spectrum described by the number
24000 the maximum of flux exhibits
about 160 counts/s/keV, and above 20 keV flux drops below 2 counts/s/keV. For such
a low flux it is impossible to obtain a signal to noise ratio which is
useful for fitting of any emission models. We also
did not attempt to fit hard
X-ray part of the spectrum since our model atmospheres are dark above 20
keV. Therefore, analyzing of hard X-ray spectrum for MXB 1728-34 with our
model spectra would not yield any informations regarding $T_{\rm eff}$, $\log g$
and the redshift $z$.

Characteristic features of our spectra are: line-like absorption
feature around 4.5 keV and flux excess around 6.5 keV. We note the
existence of such an excess in each analyzed spectrum but sometimes this
feature is not prominent. Line-like feature appears in some of spectra, but sometimes it decreases
to the apparent noise level. Interpretation of this feature is not clear, and fitting a model
gaussian line of central energy 4.5 keV do not improve the fit. Therefore we do not attempt
to model this line. What is important, the line appears only
when we use the response matrix version 1996, and in later versions of the
response matrix this problem was solved. This line can be interpreted as
the instrumental xenon L--edge line. 

Situation is quite different in case of the flux excess
around 6.5 keV. Approximating of this excess as a broad gaussian line
seems necessary. PCA data did not allow us for the physically
meaningful interpretation of this ``line''. Also determination of the
centroid energy corresponding to this line is impossible using the fitting
procedure. 
In case of spectra recorded by other instruments like MECS on
bo\-ard Bep\-poSAX satellite such a flux excess was interpreted as the 
fluorescent iron line produced in cold disk illuminated by hard X-ray photons
(see Di Salvo et al. 2000). Our model atmosphere do not include
processes which produced Fe$K_\alpha$ line, and we appended this line 
``by hand'' using the {\sc xspec} model of a gaussian line.

\begin{figure}[!h]
\includegraphics[scale=0.5,angle=270]{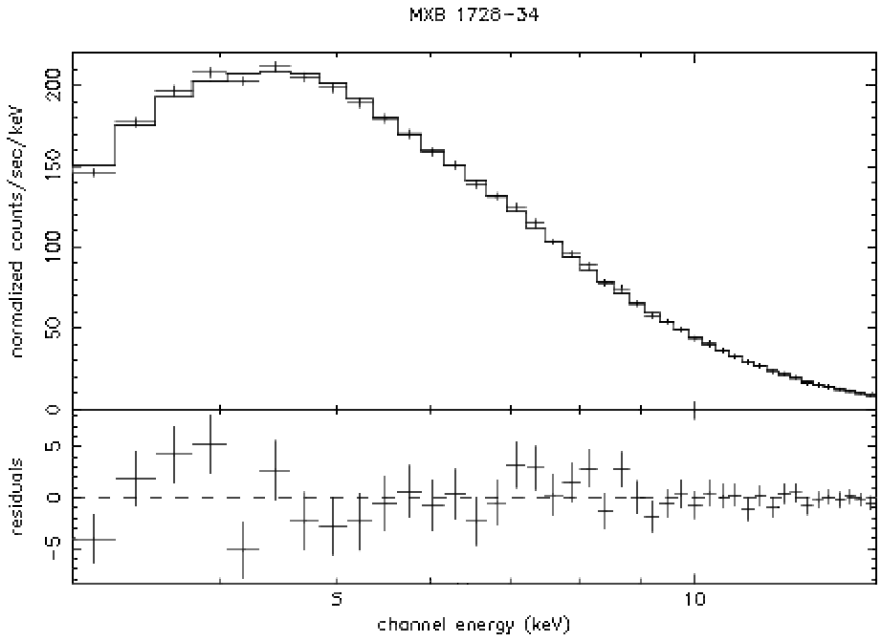}
\caption{{ \footnotesize Sample fit of the MXB 1728-34 spectrum of 96-sec
length, ObsID 10073-01-03-000. We have fitted wabs*plabs(ATM+gaussian) 
template spectra. Fit is reasonable ($\chi^2=0.812$), and the lower panel
exhibits residua.}}
\label{ATM}
\end{figure}

If we restricted our analysis and used the standard template spectra of the
{\sc xspec} software, than the best fits consist of a blackbody spectrum
plus power law component with high energy cutoff modified by interstellar
absorption of cold matter.  Moreover, if we included multicolor disk
blackbody model then the best fit yields unrealistic high temperatures
in the inner ring of the disk equal to 2.8 keV. Therefore, we replaced the above model by the 
emission from the comptonised neutron star atmosphere, corrected for low energy
absorption and a gaussian line. In our solution emission from the
accretion disk is assumed as negligibly small. X-ray burster MXB 1728-34
exhibits frequent bursts, with average time intervals of 8.4 hours
(Basinska et al. 1984). 
Therefore, its atmosphere does not cool down efficiently between bursts and
then X-rays from the colder disk do not contribute significantly. 

We performed fitting of 18 spectra from the RXTE archive using
extensive grids of theoretical spectra computed by the code {\sc ATM21}, 
The code was described in detail e.g. in Madej (1991a); Madej, Joss
\& R\'o\.za\'nska (2004); Majczyna et al. (2005), see also the
previous Section. 
    
In order to  determine the effective temperature, redshift  and
surface gravity of the neutron star we fitted observed
spectra of MXB 1728-34 with the model {\tt  wabs*plabs(ATM21+gaussian)}.
Such a fitting formula denote the sum of our theoretical spectrum plus gaussian
line modified by interstellar absorption and absorption caused by dust. 
Addition of absorption on dust is reasonable, because this burster is
located in the direction to the Galactic Center, where such type of extinction is
very high.

\section{Calculations of mass and radius}

Mass and radius of the neutron star was determined from the
values of surface gravity $\log{g}$ and gravitational redshift $z$. 
The effective temperature $T_{\rm{eff}}$ was not useful at this
step. Gravitational redshift is given by:
\begin{equation}
1+z = \left(1-\frac{2GM}{Rc^2}\right)^{-1/2}
\end{equation}
where $G$ is the gravitational constant, $M$ is the neutron star mass,
$R$ is the radius measured on the NS surface, $c$ denotes the speed of
light.
Gravitational acceleration on the NS surface equals to :
\begin{equation}
g=\frac{GM}{R^2}\left(1-\frac{2GM}{Rc^2}\right)^{-1/2}
\end{equation}
We solve Eqs. 1-2 for mass $M$ and radius $R$, and obtain the
explicit expressions:
\begin{equation}
\label{eqR}
R=\frac{z\, c^2}{2\, g}\,\frac{(2+z)}{(1+z)}
\end{equation}
\begin{equation}
\label{eqM}
M=\frac{z^2 c^4}{4gG}\, \frac{(2+z)^2}{(1+z)\, ^3}
\end{equation}

Both mass and radius of a neutron star are functions only of the surface
gravity $g$ and the gravitational redshift $z$. The effective temperature
$T_{\rm eff}$ (which is equivalent to the bolometric luminosity of an unit
area on the NS surface) does not directly influence neither $M$ nor $R$.

Moreover, our method of $M$ and $R$ determination for a neutron star
exhibits two interesting properties:

{\bf 1.} Values of both mass and radius are independent on the distance
to the source. We also do not need to estimate and compare both the
bolometric and apparent X-ray luminosities of a neutron star to obtain
$M$ and $R$.

{\bf 2.} Both mass and radius of a NS do not depend on the estimate of the
dimensionless parameter $\xi$, which is defined as the relative area of the 
NS star surface actually emitting X-rays. Our method allows one to measure
$M$ and $R$ also in cases, when only part of the surface (of unknown
value $\xi$) is visible in X-rays.

\subsection{The minimum mass of a neutron star}
For the proper determination of surface gravities and gravitational
redshifts it is necessary to estimate physically reliable minimum mass
for neutron stars (see Section 6). We turn attention
of the reader to the paper by Haensel et al. (2002), who investigated the
equation of state for dense matter and the minimum mass of neutron
stars. 

Haensel et al. (2002) predicted that minimum masses of cold nonrotating
neutron stars are of the order $M_{{\rm min}}\approx 0.09 M_\odot$, and this value
depends very weakly on the equation of state. If a neutron star rotates,
than the minimum mass $M_{{\rm min}}$ increases. Haensel et al. (2002) 
present the example that in the case of the most rapidly rotating radio 
pulsar ($P_{{\rm rot}}=1.56$ ms or frequency 641 Hz) the minimum mass of uniformly rotating cold neutron star is in the range $0.54 - 0.61 M_\odot$ depending on the exact equation of state.

Constraining of a NS mass was necessary to eliminate many of the fits
which we deem as physically unrealistic, since they could imply 
low NS masses even below $M_{\rm min}$. If this constraint was not 
introduced, then the averaged NS masses in MXB 1728-34 would lower than
these claimed in Section 7 of our paper. 

\section{Fitting procedure}
We have chosen total of 18 spectra from 6 different exposures of RXTE,
which correspond either to the island or banana states.  We proceeded in
the following way. For each model atmosphere of a given effective
temperature $T_{\rm{eff}}$ (on the NS surface),  logarithm of surface
gravity $\log{g}$, and some trial chemical composition (see the listing in
Sec. 2), we have fixed parameters of the line: central
energy $E_{\rm{line}}=6.2$ keV, and the width $\sigma=1.2$ keV. Redshift
$z$ was also fixed at some value in the range 0.00 -- 0.60. At the time of
fitting we iterated the following free parameters: hydrogen column density,
two parameters of the {\tt plabs} model, normalization of the ATM21 model,
and the  normalization of the line (total 5 free parameters).

We iterated the above five free parameters until the minimum of $\chi^2$
has been achieved. Such a procedure was repeated for all other
combinations of $T_{\rm{eff}}$, $\log{g}$ and $z$, and for all three
available chemical composition. Trial values of the gravitational redshift
$z$ were changed from 0.00 to 0.60 with steps of 0.01.

The actual fitting was restricted to the effective temperatures
$T_{\rm{eff}}$ in the range from $1\times 10^7$ K to
$3 \times 10^7$ K, changing with step of $ 10^6$ K. In such a way we
have obtained the table of more than 21 000 values of $\chi^2$, each of 
them corresponding to the unique set of the three fixed parameters: 
$T_{\rm{eff}}$, $\log{g}$, and $z$, for a given chemical composition. Then,
we searched for the minimum of $\chi^2$ in the three-dimensional space.

Inspection of the table of $\chi^2$ showed, that the best fitted models
with iron abundance 100 times greater than the solar value produced
significantly higher $\chi^2$ at numerous minima, than those obtained
for hydrogen-helium models. We believe that iron rich models do not
represent relevant chemical composition of the neutron star surface in 
quiescent state of MXB 1728-34. Therefore, fits to iron rich models
will neither be presented nor discussed in subsequent Sections.

Note, that in the case of a blackbody spectrum one can determine
only the quantity $(1+z) T_{\rm{eff}}$, and the value of $\log g$ remains
unknown. Separation of $T_{\rm{eff}}$ and $z$, and the determination of
$\log g$ is possible only because shapes of real stellar spectra deviate
from a blackbody. This is sometimes a marginal effect, therefore, the
redshift and the surface gravity determinations can be uncertain.

\section{Redshift and surface gravity determination}
\label{sec:loggz}
As is well known, fitting of the observed X-ray spectra with the {\sc
xspec} software usually do not produce unique results. In this research we
did not obtain a unique set of $T_{{\rm eff}}$, $\log g$ and $z$, which
would have a value of $\chi^2$ distinctly lower than remaining fits.
Instead, we always obtained numerous sets of the above fitting parameters
with $\chi^2$ very close to the minimum value.

For a given chemical composition we have selected the set of $T_{{\rm
eff}}$, $\log g$ and $z$  from thousands of fits which corresponds to the
minimum value, $\chi^2_{{\rm min}}$. We have appended other sets with
$\chi^2$ in the range $[\chi^2_{{\rm min}}, \chi^2_{{\rm min}}+\Delta_1]$,
where $\Delta_1$ represents the increase of $\chi^2$ small enough to be in
the 1-sigma confidence range. 

Sizes of various confidence ranges were estimated by Avni (1976), Lampton,
Margon \& Bowyer (1976) and Press et al. (1996). Theoretical models fitted
in this research have 5 free parameters and, typically, 38 degrees of
freedom (38 d.o.f.). We applied here values for $\Delta \chi^2$ taken from
Press et al. (1996), page 692, which give in our case 
$\Delta_1=5.89/38\,\, {\rm d.o.f.} = 0.16 $. Such a value of $\Delta_1$ was
used in the following stages of our research. 

For each of the analyzed 96 sec X-ray spectrum for MXB 1728-34 we obtained
usually few hundred sets of $T_{\rm eff}$, $\log g$ and $z$ which belong to
the $[\chi^2_{{\rm min}}, \chi^2_{{\rm min}}+\Delta_1]$ range. 

\begin{figure}[!h]
\includegraphics[scale=0.45]{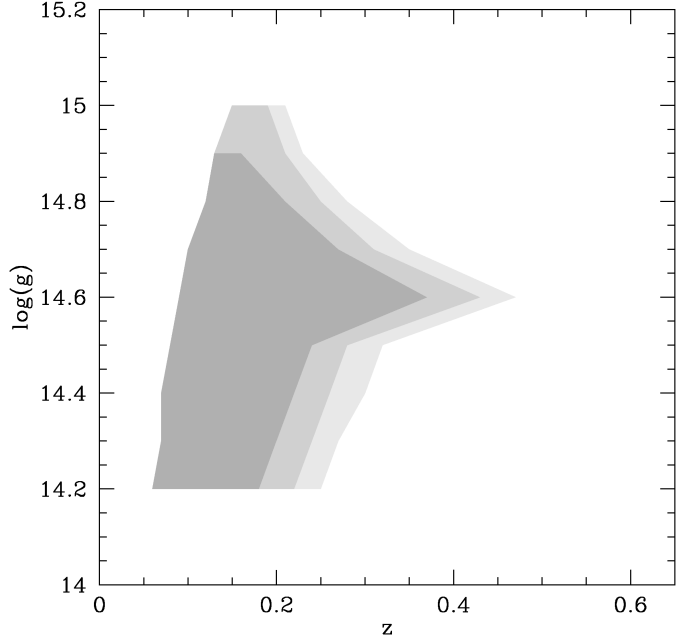}
\caption{1, 2, and 3-sigma confidence level for the fit of the model with solar
iron abundance. Here we assume that $M_{NS} \ge 0.1 M_\odot$.}
\label{konturgzfe}
\end{figure}

Shape of the confidence levels indicate that these parameters are
not correlated, see Fig. 3.

We automatically computed mass $M$ and radius $R$ from Eqs. 4 -- 5 and
rejected all sets of parameters where $M < 0.1 M_\odot$. In
such a way the number of acceptable fits was significantly reduced in order
to secure physical reliability of fits used in further considerations.
\begin{table}[!h]
\caption{The best values and 1-$\sigma$ confidence ranges for $\log g$ and $z$}
\vspace{0.5cm}
\hspace{3.2cm}
\begin{tabular}{|c|c|c|}
\hline
                &   H-He      & H-He-Fe      \\ \hline

$z_{best}$      & 0.14        & 0.22         \\
                & 0.06 - 0.21 & 0.06 - 0.41  \\
\hline 
$\log g_{best}$ & 14.6        & 14.6         \\
                & 14.2 - 14.9 & 14.2 - 14.9  \\
\hline
\end{tabular}
\label{tab:1sigma_gz}
\end{table}

Our determinations of the surface gravitational redshift can be compared
with the value of $z=0.35$ given by Cottam et al. (2002) for other neutron 
star, EXO 0748-676.

Our determinations of the surface gravity $\log g$ for the neutron
star MXB 1728-34 are in agreement with theoretical considerations by Bejger
and Haensel (2004). Both authors determined, that the maximum value of the
surface gravity $\log g =14.87$ for a neutron star build up of normal
(hadronic) matter, and $\log g =14.78$ for a strange star.

{
\newdimen\digitwidth
\setbox0=\hbox{\rm0}
\digitwidth=\wd0
\catcode`?=\active
\def?{\kern\digitwidth}

\begin{table*}[!h]
\caption{ Listing of archival X-ray spectra for MXB 1728-34 chosen for
the $T_{eff}$, $\log$,
$g$ and $z$ determination with H-He model atmospheres. \\ Fitting of 
RXTE data was done with the model wabs*plabs(ATM21+gaussian),
where $g$ is expressed in ${\rm cm\,s^{-2}}$. The column ``Spectrum''
gives the number of seconds which elapsed between the beginning of the
observation run of ID number and the beginning of the actual 96 sec time
interval.}
\small{
\vspace{0.3cm}
\hspace{-1.2cm}
\begin{tabular}{|c|c|c|c|c|c|c|c|}
\hline
Observation ID & Spectrum & State $^{\rm a}$ & log(g) & 
\,\,\,z \,\,\, &$\chi^2_\nu$ &Mass\,[$M_\odot$] & 
Radius\,[km]\\\hline
               & ?2244 &IS& 14.1 -- 14.9 & 0.05 -- 0.24 &0.846& 0.1 -- 0.8 & 1.4 -- ?8.9 \\
10073-01-02-00 & 13072 &IS& 14.2 -- 14.9 & 0.06 -- 0.20 &0.770& 0.1 -- 0.6 & 1.4 -- ?7.5 \\
\hline
               & ??768 &IS& 14.2 -- 14.9 & 0.06 -- 0.23 &0.815& 0.1 -- 0.7 & 1.4 -- ?8.4 \\
10073-01-03-000& 11904 &IS& 14.2 -- 14.9 & 0.06 -- 0.24 &0.825& 0.1 -- 0.8 & 1.4 -- ?8.9 \\
               & 13152 &IS& 14.2 -- 14.9 & 0.06 -- 0.21 &0.808& 0.1 -- 0.7 & 1.4 -- ?8.0  \\
\hline
20083-01-01-01 & ?7392 &UB& 14.2 -- 15.0 & 0.07 -- 0.24 &0.854& 0.1 -- 0.9 & 1.3 -- ?8.9 \\
               & ?7584 &UB& 14.1 -- 15.0 & 0.05 -- 0.25 &0.938& 0.1 -- 0.9 & 1.3 -- ?9.4 \\
\hline
               & ??500 &UB& 14.6 -- 14.9 & 0.09 -- 0.15 &0.720& 0.1 -- 0.3 & 1.4 -- ?3.2 \\            
               & ??700 &UB& 14.1 -- 15.0 & 0.05 -- 0.28 &0.628& 0.1 -- 1.3 & 1.3 -- 11.4 \\
               & ?1000 &UB& 14.1 -- 15.0 & 0.05 -- 0.25 &0.906& 0.1 -- 1.0 & 1.3 -- ?9.9 \\
20083-01-02-01 & ?1200 &UB& 14.6 -- 14.9 & 0.10 -- 0.17 &0.850& 0.1 -- 0.3 & 1.4 -- ?3.4 \\               
               & ?2000 &UB& 14.2 -- 15.0 & 0.07 -- 0.24 &0.750& 0.1 -- 0.9 & 1.3 -- ?9.4 \\
               & ?6000 &UB& 14.1 -- 14.9 & 0.05 -- 0.24 &0.938& 0.1 -- 0.9 & 1.4 -- ?9.4 \\
\hline
               & ?1500 &LB& 14.2 -- 14.9 & 0.06 -- 0.20 &0.576& 0.1 -- 0.6 & 1.4 -- ?7.5 \\
               & ?7300 &LB& 14.2 -- 14.9 & 0.06 -- 0.23 &0.897& 0.1 -- 0.7 & 1.4 -- ?8.5 \\
20083-01-04-00 & 12240 &LB& 14.2 -- 14.9 & 0.06 -- 0.23 &0.845& 0.1 -- 0.8 & 1.4 -- ?8.9 \\
               & 24000 &LB& 14.2 -- 14.6 & 0.06 -- 0.14 &0.754& 0.1 -- 0.3 & 1.9 -- ?4.4 \\ 
\hline
20083-01-04-01 & 24900 &LB& 14.4 -- 14.5 & 0.07 -- 0.10 &0.915& 0.1 -- 0.2 & 2.2 -- ?3.1 \\
\hline
\end{tabular}
  }
\label{tab:hhe}
\begin{list}{}{}
\item[$^{\mathrm{a}}$] Spectral states: IS -- island state, UB -- 
upper banana state, LB -- lower banana state.
\end{list}
\end{table*}
}

{
\newdimen\digitwidth
\setbox0=\hbox{\rm0}
\digitwidth=\wd0
\catcode`?=\active
\def?{\kern\digitwidth}

\begin{table*}[!h]
\caption{The same for fits to the grid of model atmospheres with solar iron abundance.}
\small{
\vspace{0.3cm}
\hspace{-1.2cm}
\begin{tabular}{|c|c|c|c|c|c|c|c|}
\hline
Observation ID & Spectrum & State &log(g) & \,\,\,
z \,\,\, &$\chi^2_\nu$ &Mass\,[$M_\odot$] & Radius\,[km]\\
\hline
               &?2244&IS& 14.2 -- 14.9 & 0.06 -- 0.37 &0.848& 0.1 -- 1.1 & 1.4 -- ?9.4 \\
10073-01-02-00 &13072&IS& 14.2 -- 14.8 & 0.06 -- 0.30 &0.772& 0.1 -- 0.8 & 1.6 -- ?6.4 \\ \hline
               &??768&IS& 14.2 -- 14.9 & 0.06 -- 0.34 &0.814& 0.1 -- 1.0 & 1.4 -- ?8.5 \\ 
10073-01-03-000&11904&IS& 14.2 -- 14.9 & 0.06 -- 0.30 &0.805& 0.1 -- 0.8 & 1.4 -- ?8.0 \\               
               &13152&IS& 14.2 -- 14.8 & 0.06 -- 0.34 &0.833& 0.1 -- 1.0 & 1.6 -- ?8.0 \\ \hline
20083-01-01-01 &?7392&UB& 14.2 -- 15.0 & 0.06 -- 0.44 &0.885& 0.1 -- 1.5 & 1.3 -- 11.4 \\               
               &?7584&UB& 14.2 -- 15.0 & 0.06 -- 0.38 &0.836& 0.1 -- 1.2 & 1.3 -- ?9.4 \\
\hline
               &??500&UB& 14.2 -- 14.9 & 0.06 -- 0.39 &0.997& 0.1 -- 1.2 & 1.4 -- ?8.9 \\
               &??700&UB& 14.2 -- 15.0 & 0.06 -- 0.49 &0.569& 0.1 -- 2.0 & 1.3 -- 14.6 \\
               &?1000&UB& 14.2 -- 15.0 & 0.06 -- 0.45 &0.968& 0.1 -- 1.5 & 1.3 -- 11.8 \\
20083-01-02-01 &?1200&UB& 14.2 -- 15.0 & 0.06 -- 0.42 &1.000& 0.1 -- 1.4 & 1.3 -- 10.4 \\
               &?2000&UB& 14.2 -- 15.0 & 0.06 -- 0.44 &0.729& 0.1 -- 1.5 & 1.3 -- 11.8 \\
               &?6000&UB& 14.2 -- 15.0 & 0.06 -- 0.43 &0.993& 0.1 -- 1.4 & 1.3 -- 10.9 \\
\hline
               &?1500&LB& 14.2 -- 14.9 & 0.06 -- 0.38 &0.552& 0.1 -- 1.2 & 1.4 -- ?8.9 \\
               &?7300&LB& 14.2 -- 14.8 & 0.06 -- 0.34 &0.932& 0.1 -- 1.0 & 1.6 -- ?6.7 \\
20083-01-04-00 &12240&LB& 14.2 -- 14.9 & 0.06 -- 0.39 &0.843& 0.1 -- 1.2 & 1.4 -- ?9.4 \\
               &24000&LB& 14.2 -- 15.0 & 0.06 -- 0.34 &0.769& 0.1 -- 1.0 & 1.6 -- ?6.7 \\ 
\hline
20083-01-04-01 &24900&LB& 14.6 -- 14.6 & 0.09 -- 0.16 &0.985& 0.1 -- 0.3 & 1.9 -- ?3.4 \\ 
\hline
\end{tabular}
  }
\label{tab:fe}
\end{table*}
}

\section{Mass and radius determination}
\label{sec:diss}

This Section presents the final results of our analysis for both assumed
chemical compositions.

Our results are explained in detail in Tables 2
and 3. Each row of both Tables presents detailed analysis for
each of the 96 sec spectra taken into account in this paper. Columns of
both Tables give: identification of the observation, name of the spectrum,
the range of $\log g$ and $z$ which are located in the 1-$\sigma$
confidence range, and the minimum $\chi^2$. The last two columns present
the final results: ranges of the NS masses and radii corresponding to the
1-$\sigma$ confidence range.

In order to properly weight the averaged 1-$\sigma$ limits of investigated
parameters, we have computed the arithmetic averages of the lower and upper
limits for $\log{g}\,$, $z\,$, $M$ and $R$ in Tables 2 and 3. Results also
are displayed in Tables 1 and 4
and they represent the best values of parameters for the
compact star in MXB 1728-34. 

The averaged values of $M$, $R$ and their 1-$\sigma$ confidence ranges are
presented in Table~4. 

\begin{table}[!h]
\caption{The best values and 1-$\sigma$ confidence ranges for $M$ and $R$}
\label{tab:1sigma_mr}
\vspace{0.3cm}
\hspace{3.2cm}
\begin{tabular}{|c|c|c|}
\hline
                            & H-He      & H-He-Fe    \\ \hline
$M_{ \rm best}\, [M_\odot]$ & 0.40      & 0.63       \\
                            & 0.1 - 0.7 & 0.1 - 1.2  \\ \hline 

$R_{\rm best}\, [{\rm km}]$ & 4.58      & 5.27       \\
                            & 1.4 - 7.7 & 1.4 - 9.1  \\
\hline
\end{tabular}
\end{table}

As is evident, both mass and radius of the neutron star in MXB 1728-34 are
very low as compared with canonical values. This is true for both H-He and
H-He-Fe chemical compositions of a neutron star atmosphere. In any event,
our $M$ and $R$ determinations are consistent with a family of currently
known equations of state for superdense matter. Our best values of mass $M$
and radius $R$ clearly suggest, that the neutron star in MXB 1728-34 is
composed of strange matter (see Figs. 4-5). 

Results presented in Tables 1 -- 4 clearly suggest, that the
canonical mass of 1.4 $M_\odot$ is rather excluded by almost all of our
detailed fits, no matter what is the exact chemical composition of model
atmospheres among the two presented in this paper. There exists only a
single 96 s spectrum in Table 3
 which contradicts this
statement, however, that fit is not really good because the redshift $z$ is
very poorly constrained. Therefore, it yields rather indefinite results on
both the mass $M$ and radius $R$. 

\begin{figure}[!h]
\includegraphics[scale=0.48]{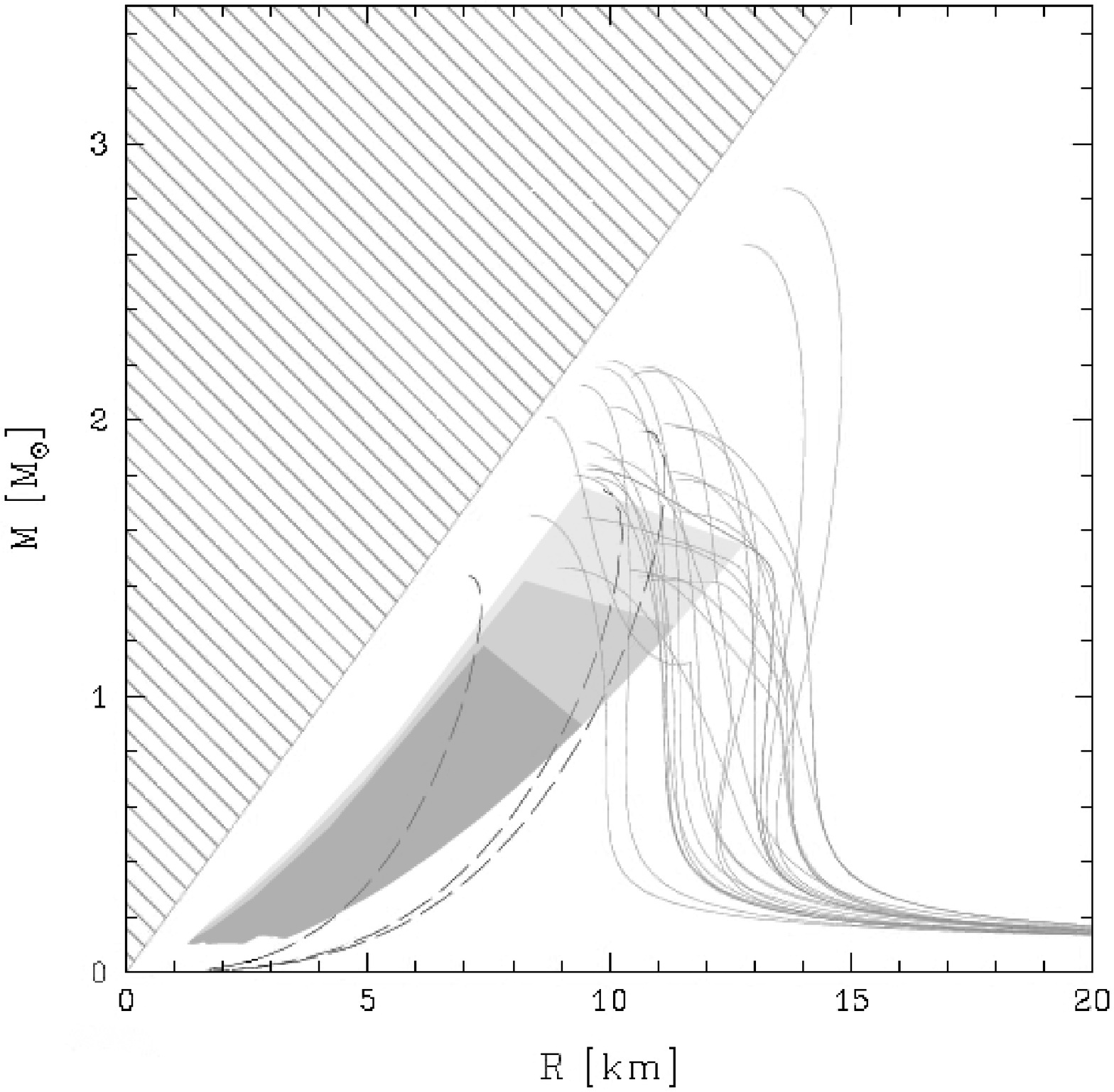}
\caption{{\footnotesize Position of MXB 1728-34 on the $M-R$ diagram 
determined from models with iron in solar proportion. We
adapted Fig. 2 from Bejger and Haensel (2002), who plotted here examples
of 30 equations of state for quark matter (dashed lines), and normal matter
or normal matter plus strange matter (solid lines). The shaded area is
excluded by General Relativity and causality condition. Our determinations
for the $M$ and $R$ confidence ranges are shown by gray contours:
dark gray area is the 1-$\sigma$ confidence level, medium -- 2-$\sigma$ and
light gray area -- 3-$\sigma$ confidence level.}}
\label{fig:EOS}
\end{figure}

\begin{figure}[!h]
\includegraphics[scale=0.48]{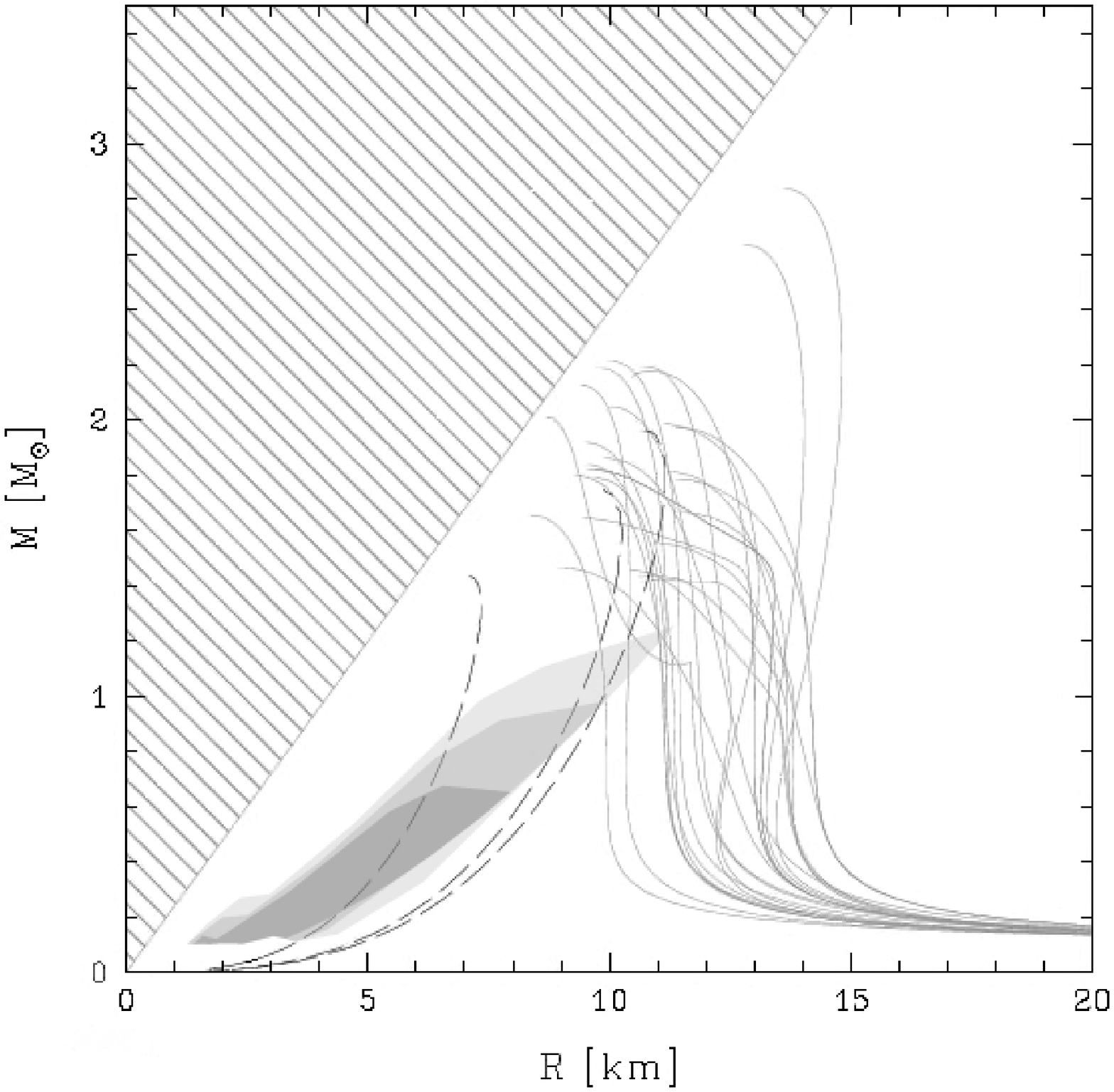}
\caption{{\footnotesize Position of MXB 1728-34 on the $M-R$ diagram 
determined from models containing hydrogen and helium in solar
proportion. We
adapted Fig. 2 from Bejger and Haensel (2002), who plotted here examples
of 30 equations of state for quark matter (dashed lines), and normal matter
or normal matter plus strange matter (solid lines). The shaded area is
excluded by General Relativity and causality condition. Our determinations
of the NS confidence ranges of the parameters are denoted by gray
areas: dark gray area is the 1-$\sigma$ confidence level, medium -- 
2-$\sigma$ and light gray area -- 3-$\sigma$ confidence level.}}
\label{fig:EOS}
\end{figure}

The radius $R=10 - 12$ km is thought to represent the canonical
value. Tables 1 and 4
show that these values are on the fringe of our 1-$\sigma$ confidence range.
In other words, centroids of fits in Figs. 1 -- 4 
show that the neutron star in MXB 1728-34 is much smaller and less
massive than the canonical NS star. 

Note, that masses M and radii R determined in this paper are partially
correlated, as is shown in
Figs. 3 -- 4. This means, that the area on the M $\times$ R plane which is
permitted by our determinations is much smaller than the area of the
rectangle of the size given by the confidence ranges from Table 4.


The above results strongly suggest, that the ``neutron'' star in MXB
1728-34 is composed of strange matter, see Figs. 3 -- 4.
We cannot exclude, however, that the NS star in this object is build up of normal
matter if we investigate the 1-$\sigma$ confidence range for both $M$ and $R$.


\section{Comments}

Few additional comments can be appended to our method and results, which
were presented in the previous Section. 

{\bf 1.} First, one can immediately determine the compactness 
of the MXB 1728-34, which is the ratio of mass (in solar masses) to radius
(in km), see Nath et al. (2002). We find that the compactness $M/R$
of the MXB 1728-34 source obtained it this paper equals to 0.087 (H-He
models), or 0.12 for H-He-Fe models.

Nath et al. (2002) discovered milisecond X-ray oscillations in the
rising part of light curves in two bursts of this source. Observations
were performed by RXTE in 1996-1997, and these X-ray oscillations were
subsequently fitted by radiation from either one or two hot spots on
the surface of a rotating neutron star. Taking into account the
relativistic deflection of emitted X-rays, they found the averaged value
of $M/R=0.121$ for MXB 1728-34. Our determinations of the $M/R$ are
consistent with their findings. 

{\bf 2.}
Recently Shaposhnikov and Titarchuk (2002) have published numerical
model for the computation of the structure of a neutron star atmosphere and
its X-ray spectrum for the expansion stage of X-ray burst. I.e. their model
is applicable to the phase when radiation forces approach gravitational
forces. On the contrary, we used model atmospheres in
hydrostatic equilibrium, therefore, our theoretical X-ray spectra represent
spectra of a neutron star with the luminosity significantly lower than
the Eddington luminosity (quiescent state).

Note, that Shaposhnikov and Titarchuk (2002) followed the strategy in which
their model spectra were approximated by a blackbody spectrum, where
possible. We have followed opposite strategy. Our theoretical X-ray spectra
of neutron star were never approximated by some models. In this way we
made use of even slight deformations of theoretical continuum spectra as
compared with the blackbody, and discrete spectral features of iron to fit
X-ray observations and extract as much informations about a neutron star as
possible (i.e. values of redshift $z$ and $\log g$ simultaneously).

{\bf 3.}
We stress here, that all our determinations of parameters for MXB 1728-34
were obtained assuming the particular model for fitting, {\tt
wabs*plabs(ATM21+gau\-ssian)}. We have chosen this model also because it
produced reasonably low $\chi^2$ for the best fits. In future we plan to
seek also for other models for the same set of X-ray spectra of MXB
1728-34, as these in Tables 2 and 3.


\section{Summary}

We present in this paper the determination of mass, radius, surface gravity
and the gravitational redshift for the neutron star in X-ray burster MXB
1728-34. For the first time such results were obtained by fitting of
archival RXTE raw spectra of this source in quiescent state to grids of
advanced models of hot neutron star theoretical X-ray spectra. The latter
were computed with the {\sc ATM21} code which was extensively described
e.g. in Madej (1991a), Madej et al. (2004), and Majczyna et al. (2005).

We have fitted 18 of 96-sec RXTE spectra to three available grids of
hot NS theoretical spectra, corresponding to different chemical composition
of models. Two grids of models yielded very good fits: the hydrogen-helium
model atmospheres with solar He abundance (hereafter the grid 1), and the
H-He-Fe grid with solar iron abundance hereafter the grid 2). We ruled out
the grid of H-He-Fe mixture with iron abundance 100 times the solar value,
since it always yielded much worse $\chi^2$ values than the first two
chemical compositions.

Both the former chemical compositions ensure fits of similar credibility.
We conclude that the best determined parameters of the neutron star in 
MXB 1728-34 are: the surface gravity is $\log g =14.6$ (grid 1) or
$14.6$ (grid 2), and the surface redshift $z = 0.14$ (1) or  $0.22$ (2).
There are, however, rather large sizes of 1-$\sigma$ confidence ranges,
particularly for the surface redshift $z$, see Section 6 
These parameters imply low values of mass $M$ and radius $R$ of the
neutron star: mass $M=0.40\, M_\odot$ (1) or $0.63\, M_\odot$ (2),
and the radius $R=4.58$ km (1) or $5.27$ km (2).

Note, that these results were obtained
assuming the particular model for fitting, {\tt wabs*plabs(ATM+gaussian)}.

Our method of the surface gravity, gravitational redshift, mass
and radius determination for a neutron star takes into account the shape
of a NS continuum X-ray spectrum and its location on the energy-flux
diagram.
The method does not make any reference to the (poorly known)
distance to the source and its bolometric or Eddington luminosities. 
Our method of the determination of the above four (global) parameters
of the NS does not make any reference to the question whether the
whole surface of the NS or only part of it is visible in X-ray satellite
detectors.

Our results imply that the NS star in MXB 1728-34 in fact is composed of
strange matter, see Figs. 3 and 4.
Therefore, we confirm the hypothesis proposed by  Li et al. (1999), who noted that the M-R relation
for MXB 1728-34 is typical for the equation of state of strange matter. Of course, a bare quark star
can not be the source of bursts. However, a quark star with a crust and external envelope composed
of normal matter can exhibit bursts (Miralda-Escud{\'e} et al. 1990).

\Acknow
We thank Micha{\l} Bejger, Pawe{\l} Haensel, Miros{\l}aw Na\-le\-{\.z}y\-ty and
Ewa Szuszkiewicz for their helpful commends regarding our research on MXB
1728-34. This work has been supported by the Polish Committee for
Scientific Research grant No. 1 P03D 001 26. \\
This research has made use of data obtained from the High Energy
Astrophysics Science Archive Research Center (HEASARC), provided by NASA's
Goddard Space Flight Center.

\begin {references}
\refitem {Arnaud K.A.} {1996} {in Astronomical Data Analysis Software and
 Systems} {V ASP Conf. Ser.,} {101} {17} {(eds. G.H. Jacoby and J. Barnes)}
\refitem {Avni Y.} {1976} {\ApJ} {210} {642}
\refitem {Basinska E.M., Lewin, W.H.G., Sztajno, M., Cominsky, L.R., 
    \and Marshall, F.J.} {1984} {\ApJ} {281} {337}
\refitem {Bejger M. \and Haensel P.} {2002} {A\&A} {396} {917}
\refitem {Bejger M. \and Haensel P.} {2004} {A\&A} {420} {987}
\refitem {Cottam, J., Paerels, F. \and Mendez, M.} {2002} {Nature} {420} {51}
\refitem {Claret A., Goldwurm, A., Cordier, B. et al.} {1994} {\ApJ} {423} {436}
\refitem {Di Salvo T., Iaria, R., Burderi, L., \and Robra, N.R.} {2000} 
    {\ApJ} {542} {1034}
\refitem {Forman W., Tananbaum H. \and Jones C.} {1976} {\ApJ} {206} {L29}
\refitem {Foster, A.J., Ross, R.R., Fabian, A.C.} {1986} {MNRAS} {221} {409}
\refitem {Galloway D.K., Psaltis, D., Chakrabarty, D., \and Muno, M.P. }{2003}{\ApJ}{590}{999}
\refitem {Guilbert, P.W.} {1981} {MNRAS} {197} {451}
\refitem {Haensel P., Zdunik J.L. \and Douchin F,} {2002} {A\&A} {385} {301}
\refitem {Joss, P.C. \& Madej, J.}{2001}{Two Years of Science with Chandra}
{}{Washington, DC, 5-7
September, 2001}
\refitem {Kaminker A.D., Pavlov, G.G., Shibanov, Y.A., Kurt, V.G., Smirnov, A.S., Shamolin, V.M., Kopaeva, I.F., \and Sheffer, E.K.} {1989} {A\&A} {220} {117}
\refitem {Lampton M., Margon B. \& Bowyer S.} {1976} {\ApJ} {208} {177}
\refitem {Margon B., Lampton M. Bowyer S., Cruddace R.} {1975} {\ApJ} {197} {25}
\refitem {Lewin W.H.G., Clark G., \and Doty J.} {1976} {IAU Circ.} {2922} {}
\refitem {Li X.-D., Ray, S., Dey, J., Dey, M., \and Bombaci, I.} {1999} {\ApJ} {527} {L51}
\refitem {Madej J.} {1974} {\Acta} {24} {327}
\refitem {Madej J.} {1991a} {\ApJ} {376} {161}
\refitem {Madej J.} {1991b} {\Acta} {41} {73}
\refitem {Madej J., Joss P.C., \and R{\'o}{\.z}a{\'n}ska A.} {2004} {\ApJ} {602} {904}
\refitem {Majczyna, A., Madej, J., Joss, P.C., R{\'o}{\.z}a{\'n}ska, A.} {2002} {High Resolution X-ray Spectroscopy with XMM-Newton and Chandra,0
    Mullard Space Science Lab., University College London, U.K.,
    24-25 October} {2002} {p. E23}
\refitem {Majczyna, A., Madej, J., Joss, P.C. \and R{\'o}{\.z}a{\'n}ska, A.} {2005} {A\&A} {430} {643}
\refitem {Mart{\'\i} J., Mirabel, I.F., Rodr{\'\i}guez, L.F., \and 
    Chaty, S.} {1998} {A\&A} {332} {L45}
\refitem {Mihalas, D.} {1978} {Stellar Atmospheres, W.H. Freeman \&
    Co., San Francisco} {} {}
\refitem {Miralda-Escud{\'e} J., Haensel P., Paczy{\'n}ski B.}{1990} {\ApJ} {362} {572}
\refitem {Nath, N.R., Strohmayer, T.E., Swank, J.H.} {2002} {\ApJ} {564} {353}
\refitem {Osherovich V. \& Titarchuk L.} {1999} {\ApJ} {522} {L113}
\refitem {Press W.H., Vetterling W.T., Teukolsky S.A., Flannery 
B.P.} {1996} {``Numerical Recipes. Second Edition.'' (Cambrige Univ. Press)} {} {}
\refitem {Shaposhnikov, N. \& Titarchuk, L.} {2002} {\ApJ} {567} {1077}
\refitem {Shaposhnikov, N., Titarchuk, L. \& Haberl, F.} {2003} {\ApJ} {593}{L35}
\refitem {Strohmayer T.E., Zhang W., Swank J.H.} {1997} {\ApJ} {487} {L77}
\refitem {Titarchuk L.} {1994} {\ApJ} {434} {570}
\refitem {Titarchuk L. \& Osherovich V.} {1999} {\ApJ} {518} {L95}
\refitem {Titarchuk L. \& Shaposhnikov N.} {2002} {\ApJ} {570} {L25}
\refitem {van Paradijs, J.} {1978} {Nature} {274} {650}
\refitem {van Paradijs, J.} {1979} {ApJ} {234} {609} 
\end {references}
\end{document}